\documentclass[aps,prd,preprint,superscriptaddress,amsmath,amssymb,showpacs]{revtex4-1}
\usepackage{dcolumn}
\usepackage{graphicx}
\usepackage{float}
\usepackage{physics}
\usepackage[colorlinks=true,allcolors=blue]{hyperref}
\usepackage{mathrsfs}
\usepackage{inputenc}

\UseRawInputEncoding
\begin{document}
\title{Holographic Schwinger effect in the dynamical AdS/QCD model}

\author{Zhou-Run Zhu}
\email{zhuzhourun@mails.ccnu.edu.cn}
\affiliation{Institute of Particle Physics and Key Laboratory of Quark and Lepton Physics (MOS), Central China Normal University,
Wuhan 430079, China}

\author{Yang-Kang Liu}
\email{ykl@mails.ccnu.edu.cn}
\affiliation{Institute of Particle Physics and Key Laboratory of Quark and Lepton Physics (MOS), Central China Normal University,
Wuhan 430079, China}

\author{Defu Hou }
\thanks{Corresponding author}
\email{houdf@mail.ccnu.edu.cn}
\affiliation{Institute of Particle Physics and Key Laboratory of Quark and Lepton Physics (MOS), Central China Normal University,
Wuhan 430079, China}

\date{\today}

\begin{abstract}
In this paper, we discuss the potential analysis of the holographic Schwinger effect in the bottom-up AdS/QCD model. We study the effect of the magnetic field on the critical field and total potential in finite chemical potential case. By evaluating the critical electric field from the DBI action, one can observe that magnetic field decreases critical electric field $E_c$. From the results of potential analysis, we find the magnetic field reduces the potential barrier and favor the Schwinger effect which agrees with the results of the critical electric field. Moreover, the Schwinger effect is more obvious when pairs are parallel to the magnetic field than that in perpendicular case in this Einstein-Maxwell-dilaton model.
\end{abstract}

\maketitle

\section{Introduction}\label{sec:01_intro}
The virtual electron-positron pairs can be real in the presence of a strong external electric-magnetic field. This non-perturbative phenomenon is described as Schwinger effect\cite{Schwinger:1951nm} in quantum electrodynamic (QED). Schwinger effect is ubiquitous in quantum field theories (QFT) which coupled to a $U(1)$ gauge field and can be seen as an universal nature of the vacuum instability when strong external fields exit. The produced particles could be scalar particles and quark-antiquark pairs.

The production rate of particle pairs was generalized to arbitrary-coupling but weak-field in \cite{Affleck:1981ag}. However, the critical electric field fails to satisfy the weak-field condition in this case. In string theory, the critical electric field $E_c$ is proportional to the string tension\cite{Fradkin:1985ys,Bachas:1992bh}. The quark and antiquark pair locate at the endpoints of string. When the string tension is lower than the strength of the electric field, the string becomes unstable and the pair production is not suppressed. At the critical electric field $E_c$, the Dirac-Born-Infeld (DBI) action becomes zero.

Using the the AdS/CFT correspondence\cite{Witten:1998qj,Gubser:1998bc,Maldacena:1997re}, the duality between the string theory in the AdS spacetime and the gauge theory on the boundary, \cite{Semenoff:2011ng} proposed a holographic perspective to deal with the Schwinger effect. The gauge group $SU(N+1)$ spontaneously breaks to $SU(N)\times U(1)$ by Higgs mechanism in the holographic step. This realize the $\mathcal{N} = 4$ super-Yang Mills (SYM) theory coupled with an $U(1)$ gauge field. Then one can obtain the potential of the quark pairs from the vacuum expectation value of a rectangular Wilson loop as pointed in \cite{Maldacena:1998im,Rey:1998ik}. The critical electric field calculated in \cite{Semenoff:2011ng} is consistent with the result from the DBI action.

The potential analysis of holographic Schwinger effect can be seen as a tunneling phenomenon. Inspired by this, \cite{Sato:2013dwa,Sato:2013hyw} performed the holographic Schwinger effect in this method. It was found that the potential barrier will vanish and the Schwinger effect occurs when the external electric field $E$ is equal or greater than the critical field $E_c$. The production rate of pairs has been discussed in \cite{Sato:2013iua,Kawai:2013xya,Zhang:2015bha,Ghodrati:2015rta,Qu:2016vqk}. \cite{Hashimoto:2014yya} has studied the production of the pairs in the non-supersymmetric QCD background. \cite{Fadafan:2015iwa} discussed the holographic Schwinger effect in the Lifshitz and hyperscaling violation theories.  \cite{Hashimoto:2013mua,Hashimoto:2014dza} performed the instability of the vacuum and magnetic in the $\mathcal{N} = 2$ supersymmetric QCD. \cite{Bolognesi:2012gr,Sato:2013pxa} considered the effects of the electric and magnetic field. \cite{Zhu:2019igg} consider the effect of magnetic field on the geometry of background and discussed the Schwinger effect when external electric field perpendicular to the magnetic field. Other illuminating work can be seen in \cite{Dietrich:2014ala,Zhang:2018hfd,Dehghani:2015gtd,Fischler:2014ama,Ambjorn:2011wz,Chakrabortty:2014kma,Zhang:2020noe,Li:2018lsl,Zhang:2018oie,Zhou:2021nbp}.

The heavy ion collision experiments in RHIC and LHC provide the possibility to observe the Schwinger effect since a strong electro-magnetic field is generated. The quark-antiquark pair or electron-positron pair may be produced in this extreme environment. Moreover, a strong magnetic field has been observed in the early stages of noncentral relativistic heavy ion collisions\cite{Skokov:2009qp,Voronyuk:2011jd,Bzdak:2011yy,Deng:2012pc,Huang:2015oca}. Although the magnetic field decreases rapidly after the heavy ion collisions, it is still very strong at the initial formation of the QGP \cite{Tuchin:2013apa,McLerran:2013hla}. The magnetic field is expected to affect the plasma evolution and the charge dynamics in the plasma. This expectation leads to the further study of QCD in the magnetized system. In the further research, the magnetic field has significant influences on the QCD matter and phase diagram \cite{DElia:2010abb,Bali:2011qj,Miransky:2015ava}. The interesting properties of the magnetic field drew much attention in different subjects \cite{DHoker:2009ixq,Dudal:2015wfn,Zhang:2020ben,Shi:2019wzi,Feng:2019boe,Zhang:2020efz,Chen:2021gop,Arefeva:2020bjk}.

Since the magnetic field has an influence on the background geometry, studying the holographic Schwinger effect in the magnetized background may give different inspiration on the production rate of particle pairs. Moreover, the chemical potential serves as a fundamental parameter of QGP. In this paper, we discuss the effects of the magnetic field on holographic Schwinger effect with finite chemical potential in the bottom-up dynamical AdS/QCD model of \cite{Bohra:2019ebj}. It should be mentioned that \cite{Bohra:2019ebj} found the QCD string tension in the perpendicular case increases with magnetic field while the string tension decreases with magnetic field in parallel case. Moreover, \cite{Zhou:2020ssi} used this model to study the thermodynamics of heavy quarkonium and found the magnetic field in parallel case has a large influence on the free energy. The results of \cite{Bohra:2019ebj,Zhou:2020ssi} are consistent with lattice results \cite{Bonati:2016kxj} which implies this AdS/QCD model is more appropriate to describe QCD physics. From the above, we will use this reliable bottom-up AdS/QCD model to study the production rate of the real particle pairs when external electric field perpendicular and parallel to the magnetic field respectively. From the results, the magnetic field reduces critical electric field and potential barrier in finite chemical potential case. Moreover, the Schwinger effect is more obvious when pairs are parallel to the magnetic field than that in perpendicular case.

The organization of this paper is as follows. In Sec.~\ref{sec:02}, we introduce the background geometry of the AdS/QCD model. In Sec.~\ref{sec:03}, we investigate the the effect of magnetic field on the Schwinger effect with finite chemical potential. The conclusion and discussion are given in Sec.~\ref{sec:04}.
\section{Background geometry}\label{sec:02}

In this section, we review the bottom-up dynamical AdS/QCD model with a background magnetic field which put forward in \cite{Bohra:2019ebj}. The action with two Maxwell fields is
\begin{equation}
\begin{split}
\label{eq1}
 \ S = -\frac{1}{16\pi G_5 }\int d^5 x \sqrt{- g}[R-\frac{f_{1}(\phi)}{4}F_{(1)MN}F^{MN}-\frac{f_{2}(\phi)}{4}F_{(2)MN}F^{MN}-\frac{1}{2}\partial_M \phi \partial^M \phi-V(\phi) ],
 \end{split}
\end{equation}
where $\phi$ denotes the dilaton field and $V(\phi)$ represents the potential of $\phi$. $F_{(1)MN}$ and $F_{(2)MN}$ denote the field strength tensors of the two U(1) gauge fields. $f_{1}(\phi)$ and $f_{2}(\phi)$ denote the gauge kinetic functions.

In the string frame, the metric solution of this Einstein-Maxwell-dilaton model \cite{Bohra:2019ebj} with magnetic field and chemical potential can be written as
\begin{equation}
\label{eq2}
\ ds^{2}=\frac{L^2 e^{2A_s (z)}}{z^2}[-g(z)dt^2+dx_{1}^{2}+e^{B^2 z^2}(dx_{2}^{2}+dx_{3}^{2})+\frac{dz^{2}}{g(z)}],
\end{equation}
where $L$ denotes the AdS radius and $A_s (z)=A(z)+ \sqrt{\frac{1}{6}} \phi(z)$. The expressions of $A(z)$ and $\phi(z)$ are
\begin{equation}
\label{eq3}
A(z)=-a z^2
\end{equation}
and
\begin{equation}
\label{eq4}
\begin{split}
 \phi(z)&= \frac{(9a - B^2)log(\sqrt{6a^2 - B^4}\sqrt{6a^2 z^2+9a-B^4z^2 -B^2}+6a^2 z -B^4 z)}{\sqrt{6a^2 - B^4}}\\
 &+z\sqrt{6a^2 z^2+9a-B^4 z^2 -B^2}-\frac{(9a - B^2)log(\sqrt{9a - B^2}\sqrt{6a^2 - B^4})}{\sqrt{6a^2 - B^4}}.
  \end{split}
\end{equation}

The background magnetic field $B$ is introduced by employing the second gauge field with $F_{(2)MN} = Bdx_2 \wedge dx_3$. The direction of $B$ is in $x_1$-direction in this metric and magnetic field breaks the rotation symmetry. In order to match with the lattice QCD deconfinement temperature at $B=0$ \cite{Dudal:2017max}, one can take $a= 0.15\ GeV^2$. It should note that $\phi(z)$ can make sense in the condition $B^4 < 6a^2$. In following calculations, we will set $B < 0.606\ GeV$ which the condition is satisfied. It should be mentioned that this magnetic field B in the metric is 5-dimensional (mass dimension one). One can get the physical and 4-dimensional magnetic field $\mathfrak{B}$ (mass dimension 2) by rescaling $\mathfrak{B}\sim \frac{B}{L}$ \cite{DHoker:2009ixq,Dudal:2015wfn}. In order to study the qualitative features of magnetic field intuitively, we use the 5-dimensional magnetic field $B$ in the calculations.

The chemical potential $\mu$ is related to the boundary condition ($A_t (0)=\mu$) and the gauge field $A_t (z)$ is
\begin{equation}
\label{eq5}
A_t (z)=\mu[1-\frac{\int^{z}_0 d\xi \frac{\xi e^{-B^2 \xi^2}}{f_1(\xi)\sqrt{S(\xi)}}}{\int^{z_h}_0 d\xi \frac{\xi e^{-B^2 \xi^2}}{f_1(\xi)\sqrt{S(\xi)}}}]=\widetilde{\mu}\int^{z_h}_{z} d\xi \frac{\xi e^{-B^2 \xi^2}}{f_1(\xi)\sqrt{S(\xi)}},
\end{equation}
with the scale factor $S(z)$ and the gauge coupling function $f_1 (z)$
\begin{equation}
 \label{eq6}
\begin{split}
 & S(z) = e^{2A(z)},\\
 & f_1 (z) = \frac{e^{-cz^2 -B^2 z^2}}{\sqrt{S(z)}}.
 \end{split}
\end{equation}

In order to match with the lowest lying heavy meson states $J/\Psi$ and $\Psi'$, one can set the parameter $c = 1.16\ GeV^2$ as discussed in \cite{Dudal:2017max,Yang:2015aia}.

The blackening function $g(z)$ is given as
\begin{equation}
 \label{eq7}
 g(z) = 1+\int^{z}_0 d\xi \xi^3 e^{-B^2 \xi^2-3A(\xi)}[K_3 + \frac{\widetilde{\mu}^2}{2c L^2}e^{c \xi^2}],
\end{equation}
with
\begin{equation}
 \label{eq8}
 K_3 = -\frac{[1+ \frac{\widetilde{\mu}^2}{2c L^2}\int^{z_h}_0 d\xi \xi^3 e^{-B^2 \xi^2-3A(\xi)+c\xi^2}]}{\int^{z_h}_0 d\xi \xi^3 e^{-B^2 \xi^2-3A(\xi)}}.
\end{equation}

The Hawking temperature in this background is
\begin{equation}
\label{eq9}
T=-\frac{z^3_h e^{-3A(z_h)-B^2 z^2_h}}{4\pi}[K_3 + \frac{\widetilde{\mu}^2}{2c L^2}e^{c z^2_h}].
\end{equation}
where $z_{h}$ denotes the horizon. From the Fig.1 of \cite{Bohra:2019ebj}, one can find that there are two black hole phases from the relations between Hawking temperature and horizon. The temperature $T$ first decreases with horizon $z_h$ and reaches a minimum temperature $T_{min}$. This branch represents the large black hole phase and is thermodynamically stable. We will take physical values of $z_h$ in this branch in our calculations. Another branch for $T$ increases with $z_h$ is small black hole phase and thermodynamically unstable.

\section{Holographic Schwinger effect in the dynamical model}\label{sec:03}

In this section, we will discuss the effect of magnetic field on the Schwinger effect with nonzero chemical potential in the dynamical background. It is reasonable to consider the particle pair is perpendicular and parallel to the magnetic field respectively, since the magnetic field breaks
the rotation symmetry. From this point of view, we perform the potential analysis with the two cases in this background.
We consider the first case: the particle pair is perpendicular to the magnetic field. In this case, the coordinates are parameterized by
\begin{equation}
\label{eq10}
\ t=\tau,\quad x_{3}=\sigma,\quad  x_{1}=x_{2}=0,\quad z=z(\sigma).
\end{equation}

One can calculate the the Lagrangian density from the Nambu-Goto action in the metric (\ref{eq2})
\begin{equation}
\label{eq11}
\mathcal{L}=\sqrt{M(z)+N(z)\dot{z}^{2}},
\end{equation}
with
\begin{equation}
 \label{eq12}
\begin{split}
 & M(z) =  (\frac{e^{2A_s(z)}}{z^2})^2 e^{B^2 z^2}g(z), \\
 & N(z) = (\frac{e^{2A_s(z)}}{z^2})^2.
 \end{split}
\end{equation}

The conserved quantity is obtained by
\begin{equation}
\label{eq13}
\ \mathcal{L}-\frac{\partial \mathcal{L}}{\partial{\dot{z}}}\dot{z}=C.
\end{equation}

By using the boundary condition
\begin{equation}
\label{eq14}
\ \dot{z}=\frac{dz}{d \sigma}=0, \quad z=z_{c} \ (z_{0}<z_{c}<z_{h}),
\end{equation}
where $z_{0}$ represents the position of the D3-brane in the bulk and $z_{c}$ denotes the tip of U-shaped string.

Then one can get
\begin{equation}
\label{eq15}
\ \dot{z}=\frac{dz}{d \sigma}=\sqrt{\frac{M^2(z)-M(z)M(z_{c})}{M(z_{c})N(z)}}.
\end{equation}

Integrating Eq.(\ref{eq15}), one can obtain the separate length $x_{\bot}$ of the particle pair when perpendicular to the magnetic field
\begin{equation}
\label{eq16}
\ x_{\bot}=2 \int^{z_c}_{z_0} dz \sqrt{\frac{ M(z_{c}) N(z)}{ M^2(z)- M(z)M(z_{c})}}.
\end{equation}

The sum of the Coulomb potential and static energy can be obtained by combining Eq.(\ref{eq11}) with Eq.(\ref{eq15})
\begin{equation}
\label{eq17}
\ V_{(CP+SE)(\perp)}= 2T_{F}\int^{z_c}_{z_0} dz \sqrt{\frac{ M(z)N(z)}{M(z)-M(z_{c})}}.
\end{equation}
where $T_F = \frac{1}{2\pi \alpha'}$ represents the string tension.

Then we discuss the pair is parallel to the magnetic field. Then the coordinates are parameterized by
\begin{equation}
\label{eq18}
\ t=\tau,\quad x_{1}=\sigma,\quad  x_{2}=x_{3}=0,\quad z=z(\sigma).
\end{equation}

In this case, the Lagrangian density is
\begin{equation}
\label{eq19}
\mathcal{L}=\sqrt{P(z)+Q(z)\dot{z}^{2}},
\end{equation}
with
\begin{equation}
 \label{eq20}
\begin{split}
 & P(z) =  (\frac{e^{2A_s(z)}}{z^2})^2 g(z), \\
 & Q(z) = (\frac{e^{2A_s(z)}}{z^2})^2.
 \end{split}
\end{equation}

By repeating above calculations, the separate length $x_{\parallel}$ of the pair is
\begin{equation}
\label{eq21}
\ x_{\parallel}=2 \int^{z_c}_{z_0} dz \sqrt{\frac{ P(z_{c}) Q(z)}{ P^2(z)- P(z)P(z_{c})}},
\end{equation}

The sum of the Coulomb potential and static energy is
\begin{equation}
\label{eq22}
\ V_{(CP+SE)(\parallel)}= 2T_{F}\int^{z_c}_{z_0} dz \sqrt{\frac{ P(z)Q(z)}{P(z)-P(z_{c})}}.
\end{equation}

\subsection{The external electric field is  perpendicular to the magnetic field}

From above discussions, one can find the magnetic field affects the separate length $x$ and $V_{(CP+SE)}$ of the pair. Then we calculate the critical field $E_c$ from the DBI action
\begin{equation}
\label{eq23}
\ S_{DBI}=-T_{D3} \int d^{4}x \sqrt{-det(G_{\mu\nu}+\mathcal{F}_{\mu\nu})},
\end{equation}
where $T_{D3}=\frac{1}{g_{s}(2\pi)^{3}\alpha'^{2}}$ represents the D3-brane tension.

Since the magnetic field in metric(\ref{eq2}) is along the $x_{1}$ direction. We consider the external electric field $E$ is perpendicular and parallel to the magnetic field respectively. We discuss the external electric field $E$ is perpendicular to the magnetic field first. In $E\perp B$ case, we turn on $E$ in $x_{3}$ direction from $\mathcal{F}_{\mu\nu}$. The critical field $E_{c(\perp)}$ can be obtained when $S_{DBI}$ vanishes

\begin{equation}
\label{eq24}
\ E_{c(\perp)}=T_{F}\frac{e^{2A_s(z_0)}}{z^2_0}\sqrt{ e^{B^2 z^2_0}g(z_0)}.
\end{equation}

When quark-antiquark pair is perpendicular to the magnetic field and $E\perp B$, the total potential $V_{tot(\perp,\perp)}$ of the pair can be obtained as
\begin{equation}
\label{eq25}
\begin{split}
\ V_{tot(\perp,\perp)}&=V_{(CP+SE)(\perp)}-E_{\perp} x_{\perp}  \\
      &=V_{(CP+SE)(\perp)}- \alpha E_{c(\perp)} x_{\perp}.
     \end{split}
\end{equation}
where $\alpha \equiv \frac{E}{E_{c}}$ is a dimensionless parameter.

When the pair is parallel to the magnetic field and $E\perp B$, the total potential $V_{tot(\parallel,\perp)}$ of the pair is
\begin{equation}
\label{eq26}
\begin{split}
\ V_{tot(\parallel,\perp)}&=V_{(CP+SE)(\parallel)}-E_{\perp} x_{\parallel}  \\
              &=V_{(CP+SE)(\parallel)}- \alpha E_{c(\perp)} x_{\parallel}.
\end{split}
\end{equation}

\subsection{The external electric field is parallel to the magnetic field}

Then we discuss the external electric field $E$ is parallel to the magnetic field. In $E\parallel B$ case, we turn on $E$ in $x_{1}$ direction and the critical field $E_{c(\parallel)}$ can be obtained when $S_{DBI}$ vanishes

\begin{equation}
\label{eq27}
\ E_{c(\parallel)}=T_{F}\frac{e^{2A_s(z_0)}}{z^2_0}\sqrt{g(z_0)}.
\end{equation}

When the pair is  perpendicular to the magnetic field and $E\parallel B$, the total potential $V_{tot(\perp,\parallel)}$ of the pair is
\begin{equation}
\label{eq28}
\begin{split}
\ V_{tot(\perp,\parallel)}&=V_{(CP+SE)(\perp)}-E_{\parallel} x_{\perp}  \\
                    &=V_{(CP+SE)(\perp)}- \alpha E_{c(\parallel)} x_{\perp}.
     \end{split}
\end{equation}

When the pair is parallel to the magnetic field and $E\parallel B$, the total potential $V_{tot(\parallel,\parallel)}$ of the pair is
\begin{equation}
\label{eq29}
\begin{split}
\ V_{tot(\parallel,\parallel)}&=V_{(CP+SE)(\parallel)}-E_{\parallel} x_{\parallel}  \\
              &=V_{(CP+SE)(\parallel)}- \alpha E_{c(\parallel)} x_{\parallel}.
\end{split}
\end{equation}

\subsection{The results of the Schwinger effect in the dynamical model}

\begin{figure}[H]
    \centering
      \setlength{\abovecaptionskip}{-0.1cm}
    \includegraphics[width=7cm]{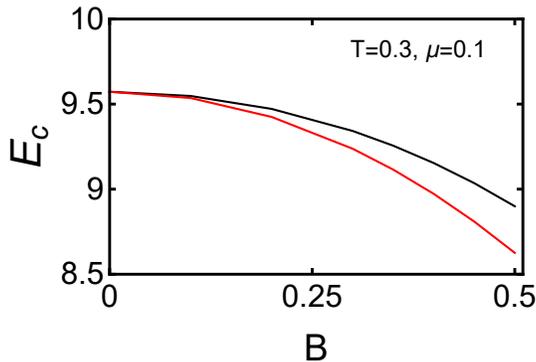}
    \caption{\label{fig1} $E_{c}$ versus $B$ when $T= 0.3\ GeV$ and $\mu = 0.1\ GeV$. The black line represents $E_{c(\perp)}$ and red line denotes $E_{c(\parallel)}$. B is in units GeV.}
\end{figure}

We will study the effect of the magnetic field on the critical field and total potential with finite chemical potential in this AdS/QCD model. We set the values of some parameters in the calculations. We take $T_F = 1$, AdS radius $L= 1$ and choose an appropriate value of $z_0$, $z_0 = 0.5$ for simplicity.

In Fig.~\ref{fig1}, we discuss the effect of the magnetic field on the $E_{c}$ when $T= 0.3\ GeV$ and $\mu = 0.1\ GeV$. We can find that $E_c$ decreases as the magnetic field increases when $E \perp B$ and $E \parallel B$, thus the Schwinger effect occurs easier when magnetic field exists. Moreover, the $E_c$ when external electric field parallel to the magnetic field is smaller than that in perpendicular case which implies the production rate of the real particle pairs is more larger when $E \parallel B$. Different from \cite{Sato:2013pxa}, the magnetic field is turned on by the second gauge field of Eq.(\ref{eq1}) in this model. The magnetic field breaks $SO(3)$ invariance in boundary and affects the geometry of background. The influences of the magnetic field on the separate length and potential barrier with nonzero chemical potential has been considered in this paper. In following figures, we will study the potential analysis in the magnetized Einstein-Maxwell-dilaton system.

\begin{figure}[H]
    \centering
      \setlength{\abovecaptionskip}{-0.1cm}
    \includegraphics[width=14cm]{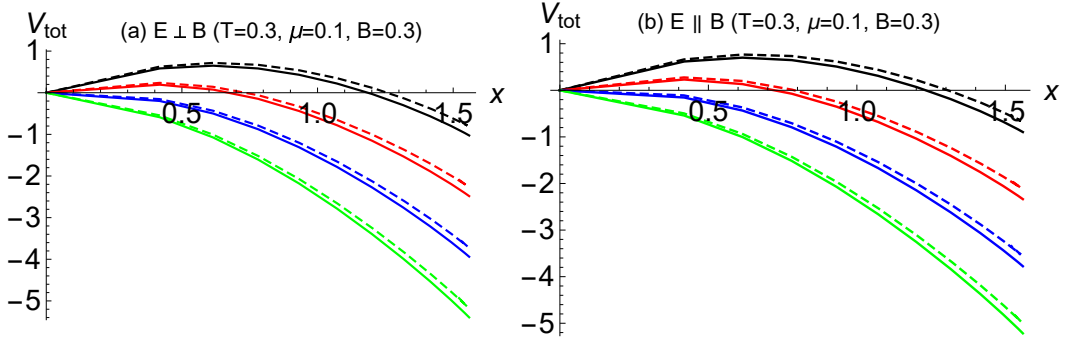}
    \caption{\label{fig2} The total potential $V_{tot}$ with respect to the separate length x with different electric field when $T = 0.3\ GeV$£¬ $\mu = 0.1\ GeV$ and $B = 0.3\ GeV$. (a) for $E \perp B$ and (b) for $E \parallel B$. The black line, red line, blue line, green line denote $\alpha(E/E_c) = 0.8,\ 0.9,\ 1.0,\ 1.1$, respectively. The solid line (dashed line) represents the pair is parallel ( perpendicular) to the magnetic field.}
\end{figure}

We plot the total potential $V_{tot}$ with respect to the separate length x with different electric field when $T = 0.3\ GeV$, $\mu = 0.1\ GeV$ and $B = 0.3\ GeV$ in Fig.~\ref{fig2}. The Schwinger effect can be related to a tunneling process. When $E < E_{c}$, the potential barrier exits and the potential barrier decreases as $E$ increases. The potential barrier vanishes when $E = E_{c}$ and the Schwinger effect is not suppressed. When $E > E_{c}$, the production of the pairs is explosive and the vacuum cannot maintain stable. The results of the potential analyses are consistent with the results of \cite{Sato:2013iua}. One also can observe that Schwinger effect is more obvious when pair is parallel to the magnetic field.

\begin{figure}[H]
    \centering
      \setlength{\abovecaptionskip}{-0.1cm}
    \includegraphics[width=14cm]{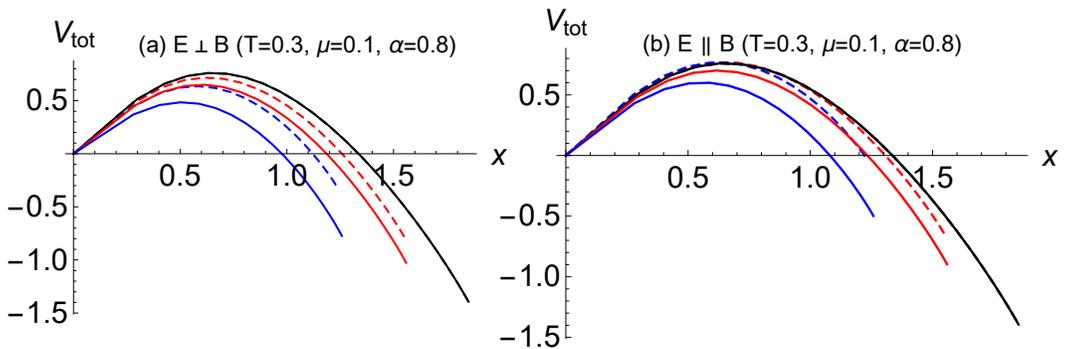}
    \caption{\label{fig3} The total potential $V_{tot}$ with respect to the separate length x with different magnetic field when $T = 0.3\ GeV$, $\mu = 0.1\ GeV$ and $\alpha = 0.8$. The black line, red line, blue line represent $B = 0,\ 0.3,\ 0.5\ GeV$ respectively. (a) for $E \perp B$ and (b) for $E \parallel B$. The solid line (dashed line) indicates the particle pair is parallel (perpendicular) to the magnetic field.}
\end{figure}

In Fig.~\ref{fig3}, we discuss the total potential $V_{tot}$ with respect to the separate length x with different magnetic field when $T = 0.3\ GeV$, $\mu = 0.1\ GeV$ and $\alpha = 0.8$. From Fig.~\ref{fig3}.(a), we can make an obvious discovery that magnetic field reduces the height and width of the potential barrier when $E \perp B$ and favor the Schwinger effect. From Fig.~\ref{fig3}.(b), the magnetic field enhance the total potential slightly at small distance $x$ when particle pair is perpendicular to the magnetic field and $E \parallel B$. But at large distance x, the magnetic field has an obvious influence on the width of the potential barrier. From this perspective, the magnetic field reduces the width of the potential barrier in large distance $x$ and enhance the Schwinger effect. Moreover, the effect of magnetic field on the potential barrier is more obvious when pair is parallel to magnetic field. It indicates Schwinger effect may occur more easily when particle pair is parallel to the magnetic field.

\begin{figure}[H]
    \centering
      \setlength{\abovecaptionskip}{-0.1cm}
    \includegraphics[width=14cm]{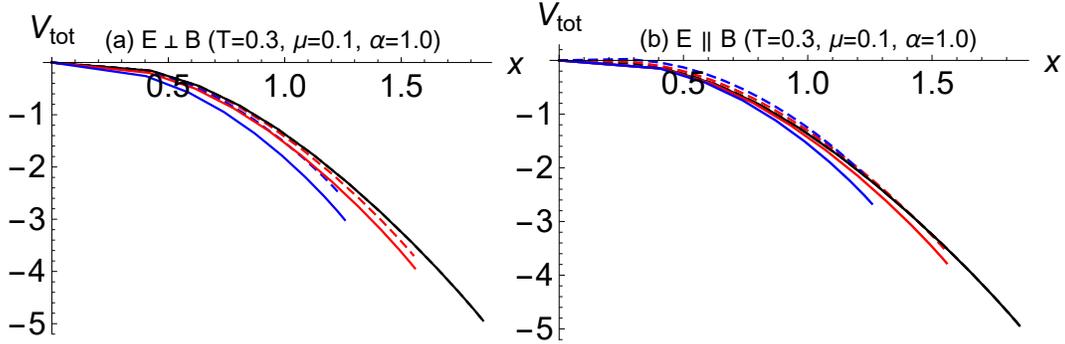}
    \caption{\label{fig4} The total potential $V_{tot}$ with respect to the separate length x with different magnetic field when $T = 0.3\ GeV$, $\mu = 0.1\ GeV$ and $\alpha = 1.0$. The black line, red line, blue line represent $B = 0,\ 0.3,\ 0.5\ GeV$, respectively. (a) for $E \perp B$ and (b) for $E \parallel B$. The solid line (dashed line) indicates the particle pair is parallel (perpendicular) to the magnetic field.}
\end{figure}

In Fig.~\ref{fig4}, we discuss the total potential $V_{tot}$ with respect to the separate length x with different magnetic field when $T = 0.3\ GeV$, $\mu = 0.1\ GeV$ and $\alpha = 1.0$. From this figure, one can find that magnetic field reduces the height and width of the potential barrier when $E \perp B$ and favor the Schwinger effect. From Fig.~\ref{fig4}.(b), the magnetic field enhances the total potential slightly when particle pair is perpendicular to the magnetic field and $E \parallel B$. However, the effect of the magnetic field on the width of the potential barrier is more prominent. The width of the potential barrier is reduced by the magnetic field and enhance the Schwinger effect.

\section{Conclusion and discussion}\label{sec:04}
A strong magnetic field was observed in the early stages of noncentral relativistic heavy ion collisions in RHIC and LHC which provides the possibility to observe the Schwinger effect. It is reasonable to consider that the magnetic field has affected the nature of the space-time geometry. In this paper, we study the Schwinger effect with finite chemical potential case in the magnetized AdS/QCD background.

In \cite{Bohra:2019ebj}, the authors found the QCD string tension decreases with magnetic field in parallel case while increases with $B$ in the perpendicular case in the magnetized AdS/QCD model. By using this model, \cite{Zhou:2020ssi} found the magnetic field in parallel case has a large influence on the free energy of heavy quarkonium. These results are consistent with lattice results \cite{Bonati:2016kxj} which implies this AdS/QCD model is more appropriate to describe QCD physics. In this paper, we study the effect of the magnetic field on the critical field and total potential with finite chemical potential case in this reliable holographic QCD model.

From the results, we find the magnetic field and chemical potential affect the critical field and the total potential of the particle pairs. One can find that the magnetic field reduces the potential barrier with nonzero chemical potential, thus favor the Schwinger effect. In the calculations of the critical electric field from the DBI action, we found magnetic field decreases $E_c$ which agrees with the results of the potential analysis. The critical field $E_c$ when external electric field parallel to the magnetic field is smaller than that in perpendicular case which implies the production rate of the real particle pairs is more larger when $E \parallel B$. Moreover, Schwinger effect is more obvious when pair is parallel to the magnetic field than that in perpendicular case in the dynamical AdS/QCD background. It implies the production rate of the real particle pairs when parallel to the magnetic field may be larger than that in perpendicular case.

We expect that this research of non-trivial magnetic field on schwinger effect in the bottom-up AdS/QCD background could provide some different afflatus in the particle pairs production. It is also interesting to study the schwinger effect under rotation from holography. One should extend the background to the rotating case \cite{Chen:2020ath}. We expect that we could report this work in the future.

\section*{Acknowledgments}

Defu Hou is in part supported by the NSFC Grant Nos. 11735007, 11890711. Zhou-Run Zhu is supported by the Fundamental Research Funds for the Central Universities (Innovation Funding Projects) 2020CXZZ100.


\begin{thebibliography}{58}

\bibitem{Schwinger:1951nm}
  J.~S.~Schwinger,
  Phys.\ Rev.\  {\bf 82}, 664 (1951).
  doi:10.1103/PhysRev.82.664

\bibitem{Affleck:1981ag}
  I.~K.~Affleck and N.~S.~Manton,
  Nucl.\ Phys.\ B {\bf 194}, 38 (1982).
  doi:10.1016/0550-3213(82)90511-9

\bibitem{Fradkin:1985ys}
  E.~S.~Fradkin and A.~A.~Tseytlin,
  Nucl.\ Phys.\ B {\bf 261}, 1 (1985)
  Erratum: [Nucl.\ Phys.\ B {\bf 269}, 745 (1986)].
  doi:10.1016/0550-3213(86)90522-5, 10.1016/0550-3213(85)90559-0

\bibitem{Bachas:1992bh}
  C.~Bachas and M.~Porrati,
  Phys.\ Lett.\ B {\bf 296}, 77 (1992)
  doi:10.1016/0370-2693(92)90806-F
  [hep-th/9209032].

\bibitem{Witten:1998qj}
  E.~Witten,
  Adv.\ Theor.\ Math.\ Phys.\  {\bf 2}, 253 (1998)
  doi:10.4310/ATMP.1998.v2.n2.a2
  [hep-th/9802150].

\bibitem{Gubser:1998bc}
  S.~S.~Gubser, I.~R.~Klebanov and A.~M.~Polyakov,
  Phys.\ Lett.\ B {\bf 428}, 105 (1998)
  doi:10.1016/S0370-2693(98)00377-3
  [hep-th/9802109].

\bibitem{Maldacena:1997re}
  J.~M.~Maldacena,
  Int.\ J.\ Theor.\ Phys.\  {\bf 38}, 1113 (1999)
  [Adv.\ Theor.\ Math.\ Phys.\  {\bf 2}, 231 (1998)]
  doi:10.1023/A:1026654312961, 10.4310/ATMP.1998.v2.n2.a1
  [hep-th/9711200].

\bibitem{Semenoff:2011ng}
  G.~W.~Semenoff and K.~Zarembo,
  Phys.\ Rev.\ Lett.\  {\bf 107}, 171601 (2011)
  doi:10.1103/PhysRevLett.107.171601
  [arXiv:1109.2920 [hep-th]].

\bibitem{Maldacena:1998im}
  J.~M.~Maldacena,
  Phys.\ Rev.\ Lett.\  {\bf 80}, 4859 (1998)
  doi:10.1103/PhysRevLett.80.4859
  [hep-th/9803002].

\bibitem{Rey:1998ik}
  S.~J.~Rey and J.~T.~Yee,
  Eur.\ Phys.\ J.\ C {\bf 22}, 379 (2001)
  doi:10.1007/s100520100799
  [hep-th/9803001].
\bibitem{Sato:2013dwa}
  Y.~Sato and K.~Yoshida,
  JHEP {\bf 1309}, 134 (2013)
  doi:10.1007/JHEP09(2013)134
  [arXiv:1306.5512 [hep-th]].

\bibitem{Sato:2013hyw}
  Y.~Sato and K.~Yoshida,
  JHEP {\bf 1312}, 051 (2013)
  doi:10.1007/JHEP12(2013)051
  [arXiv:1309.4629 [hep-th]].

\bibitem{Sato:2013iua}
  Y.~Sato and K.~Yoshida,
  JHEP {\bf 1308}, 002 (2013)
  doi:10.1007/JHEP08(2013)002
  [arXiv:1304.7917 [hep-th]].

\bibitem{Kawai:2013xya}
  D.~Kawai, Y.~Sato and K.~Yoshida,
  Phys.\ Rev.\ D {\bf 89}, no. 10, 101901 (2014)
  doi:10.1103/PhysRevD.89.101901
  [arXiv:1312.4341 [hep-th]].

\bibitem{Ghodrati:2015rta}
  M.~Ghodrati,
  Phys.\ Rev.\ D {\bf 92}, no. 6, 065015 (2015)
  doi:10.1103/PhysRevD.92.065015
  [arXiv:1506.08557 [hep-th]].

\bibitem{Zhang:2015bha}
  S.~J.~Zhang and E.~Abdalla,
  Gen.\ Rel.\ Grav.\  {\bf 48}, no. 5, 60 (2016)
  doi:10.1007/s10714-016-2056-z
  [arXiv:1508.03364 [hep-th]].


\bibitem{Qu:2016vqk}
  F.~Qu and D.~f.~Zeng,
  Phys.\ Rev.\ D {\bf 94}, no. 12, 126004 (2016)
  doi:10.1103/PhysRevD.94.126004
  [arXiv:1611.04009 [hep-th]].

\bibitem{Hashimoto:2014yya}
  K.~Hashimoto, T.~Oka and A.~Sonoda,
  JHEP {\bf 1506}, 001 (2015)
  doi:10.1007/JHEP06(2015)001
  [arXiv:1412.4254 [hep-th]].

\bibitem{Fadafan:2015iwa}
  K.~Bitaghsir Fadafan and F.~Saiedi,
  Eur.\ Phys.\ J.\ C {\bf 75}, no. 12, 612 (2015)
  doi:10.1140/epjc/s10052-015-3839-1
  [arXiv:1504.02432 [hep-th]].

\bibitem{Hashimoto:2013mua}
  K.~Hashimoto and T.~Oka,
  JHEP {\bf 1310}, 116 (2013)
  doi:10.1007/JHEP10(2013)116
  [arXiv:1307.7423 [hep-th]].

\bibitem{Hashimoto:2014dza}
  K.~Hashimoto, T.~Oka and A.~Sonoda,
  JHEP {\bf 1406}, 085 (2014)
  doi:10.1007/JHEP06(2014)085
  [arXiv:1403.6336 [hep-th]].

\bibitem{Bolognesi:2012gr}
  S.~Bolognesi, F.~Kiefer and E.~Rabinovici,
  JHEP {\bf 1301}, 174 (2013)
  doi:10.1007/JHEP01(2013)174
  [arXiv:1210.4170 [hep-th]].

\bibitem{Sato:2013pxa}
  Y.~Sato and K.~Yoshida,
  JHEP {\bf 1304}, 111 (2013)
  doi:10.1007/JHEP04(2013)111
  [arXiv:1303.0112 [hep-th]].

\bibitem{Zhu:2019igg}
Z.~R.~Zhu, D.~f.~Hou and X.~Chen,
Eur. Phys. J. C \textbf{80} (2020) no.6, 550
doi:10.1140/epjc/s10052-020-8110-8
[arXiv:1912.05806 [hep-ph]].

\bibitem{Dietrich:2014ala}
  D.~D.~Dietrich,
  Phys.\ Rev.\ D {\bf 90}, no. 4, 045024 (2014)
  doi:10.1103/PhysRevD.90.045024
  [arXiv:1405.0487 [hep-ph]].

\bibitem{Zhang:2018hfd}
  L.~Zhang, D.~F.~Hou and J.~Li,
  Eur.\ Phys.\ J.\ A {\bf 54}, no. 6, 94 (2018).
  doi:10.1140/epja/i2018-12524-4

\bibitem{Dehghani:2015gtd}
  L.~Shahkarami, M.~Dehghani and P.~Dehghani,
  Phys.\ Rev.\ D {\bf 97}, no. 4, 046013 (2018)
  doi:10.1103/PhysRevD.97.046013
  [arXiv:1511.07986 [hep-th]].

\bibitem{Fischler:2014ama}
  W.~Fischler, P.~H.~Nguyen, J.~F.~Pedraza and W.~Tangarife,
  Phys.\ Rev.\ D {\bf 91}, no. 8, 086015 (2015)
  doi:10.1103/PhysRevD.91.086015
  [arXiv:1411.1787 [hep-th]].




\bibitem{Ambjorn:2011wz}
  J.~Ambjorn and Y.~Makeenko,
  Phys.\ Rev.\ D {\bf 85}, 061901 (2012)
  doi:10.1103/PhysRevD.85.061901
  [arXiv:1112.5606 [hep-th]].


\bibitem{Chakrabortty:2014kma}
  S.~Chakrabortty and B.~Sathiapalan,
  Nucl.\ Phys.\ B {\bf 890}, 241 (2014)
  doi:10.1016/j.nuclphysb.2014.11.010
  [arXiv:1409.1383 [hep-th]].

\bibitem{Zhang:2020noe}
Z.~Q.~Zhang, X.~Zhu and D.~F.~Hou,
Phys. Rev. D \textbf{101} (2020) no.2, 026017
doi:10.1103/PhysRevD.101.026017
[arXiv:2001.02321 [hep-th]].


\bibitem{Li:2018lsl}
F.~Li, Z.~Q.~Zhang and G.~Chen,
Chin. Phys. C \textbf{42} (2018) no.12, 123109
doi:10.1088/1674-1137/42/12/123109
[arXiv:1809.10898 [hep-th]].

\bibitem{Zhang:2018oie}
Z.~q.~Zhang,
Nucl. Phys. B \textbf{935} (2018), 377-387
doi:10.1016/j.nuclphysb.2018.08.020

\bibitem{Zhou:2021nbp}
J.~Zhou and J.~Ping,
[arXiv:2101.08105 [hep-th]].

\bibitem{Skokov:2009qp}
  V.~Skokov, A.~Y.~Illarionov and V.~Toneev,
  Int.\ J.\ Mod.\ Phys.\ A {\bf 24}, 5925 (2009)
  doi:10.1142/S0217751X09047570
  [arXiv:0907.1396 [nucl-th]].

\bibitem{Voronyuk:2011jd}
  V.~Voronyuk, V.~D.~Toneev, W.~Cassing, E.~L.~Bratkovskaya, V.~P.~Konchakovski and S.~A.~Voloshin,
  Phys.\ Rev.\ C {\bf 83}, 054911 (2011)
  doi:10.1103/PhysRevC.83.054911
  [arXiv:1103.4239 [nucl-th]].

\bibitem{Bzdak:2011yy}
  A.~Bzdak and V.~Skokov,
  Phys.\ Lett.\ B {\bf 710}, 171 (2012)
  doi:10.1016/j.physletb.2012.02.065
  [arXiv:1111.1949 [hep-ph]].


\bibitem{Deng:2012pc}
  W.~T.~Deng and X.~G.~Huang,
  Phys.\ Rev.\ C {\bf 85}, 044907 (2012)
  doi:10.1103/PhysRevC.85.044907
  [arXiv:1201.5108 [nucl-th]].

\bibitem{Huang:2015oca}
  X.~G.~Huang,
  Rept.\ Prog.\ Phys.\  {\bf 79}, no. 7, 076302 (2016)
  doi:10.1088/0034-4885/79/7/076302
  [arXiv:1509.04073 [nucl-th]].

\bibitem{Tuchin:2013apa}
K.~Tuchin,
Phys. Rev. C \textbf{88} (2013) no.2, 024911
doi:10.1103/PhysRevC.88.024911
[arXiv:1305.5806 [hep-ph]].

\bibitem{McLerran:2013hla}
L.~McLerran and V.~Skokov,
Nucl. Phys. A \textbf{929} (2014), 184-190
doi:10.1016/j.nuclphysa.2014.05.008
[arXiv:1305.0774 [hep-ph]].

\bibitem{DElia:2010abb}
  M.~D'Elia, S.~Mukherjee and F.~Sanfilippo,
  Phys.\ Rev.\ D {\bf 82}, 051501 (2010)
  doi:10.1103/PhysRevD.82.051501
  [arXiv:1005.5365 [hep-lat]].



\bibitem{Bali:2011qj}
  G.~S.~Bali, F.~Bruckmann, G.~Endrodi, Z.~Fodor, S.~D.~Katz, S.~Krieg, A.~Schafer and K.~K.~Szabo,
  JHEP {\bf 1202}, 044 (2012)
  doi:10.1007/JHEP02(2012)044
  [arXiv:1111.4956 [hep-lat]].

\bibitem{Miransky:2015ava}
  V.~A.~Miransky and I.~A.~Shovkovy,
  Phys.\ Rept.\  {\bf 576}, 1 (2015)
  doi:10.1016/j.physrep.2015.02.003
  [arXiv:1503.00732 [hep-ph]].


\bibitem{DHoker:2009ixq}
E.~D'Hoker and P.~Kraus,
JHEP \textbf{03} (2010), 095
doi:10.1007/JHEP03(2010)095
[arXiv:0911.4518 [hep-th]].

\bibitem{Dudal:2015wfn}
D.~Dudal, D.~R.~Granado and T.~G.~Mertens,
Phys. Rev. D \textbf{93} (2016) no.12, 125004
doi:10.1103/PhysRevD.93.125004
[arXiv:1511.04042 [hep-th]].

\bibitem{Zhang:2020ben}
C.~Zhang, R.~H.~Fang, J.~H.~Gao and D.~F.~Hou,
Phys. Rev. D \textbf{102} (2020) no.5, 056004
doi:10.1103/PhysRevD.102.056004
[arXiv:2005.08512 [hep-th]].

\bibitem{Shi:2019wzi}
S.~Shi, H.~Zhang, D.~Hou and J.~Liao,
Phys. Rev. Lett. \textbf{125} (2020), 242301
doi:10.1103/PhysRevLett.125.242301
[arXiv:1910.14010 [nucl-th]].

\bibitem{Feng:2019boe}
S.~Q.~Feng, Y.~Q.~Zhao and X.~Chen,
Phys. Rev. D \textbf{101} (2020) no.2, 026023
doi:10.1103/PhysRevD.101.026023
[arXiv:1910.05668 [hep-ph]].

\bibitem{Zhang:2020efz}
H.~X.~Zhang, J.~W.~Kang and B.~W.~Zhang,
Eur. Phys. J. C \textbf{81} (2021) no.7, 623
doi:10.1140/epjc/s10052-021-09409-w
[arXiv:2004.08767 [hep-ph]].

\bibitem{Chen:2021gop}
X.~Chen, L.~Zhang and D.~Hou,
[arXiv:2108.03840 [hep-ph]].

\bibitem{Arefeva:2020bjk}
I.~Y.~Aref'eva, K.~Rannu and P.~Slepov,
[arXiv:2012.05758 [hep-th]].

\bibitem{Bohra:2019ebj}
  H.~Bohra, D.~Dudal, A.~Hajilou and S.~Mahapatra,
  Phys.\ Lett.\ B {\bf 801}, 135184 (2020)
  doi:10.1016/j.physletb.2019.135184
  [arXiv:1907.01852 [hep-th]].

\bibitem{Zhou:2020ssi}
J.~Zhou, X.~Chen, Y.~Q.~Zhao and J.~Ping,
Phys. Rev. D \textbf{102} (2020) no.8, 086020
doi:10.1103/PhysRevD.102.086020
[arXiv:2006.09062 [hep-ph]].

\bibitem{Bonati:2016kxj}
C.~Bonati, M.~D'Elia, M.~Mariti, M.~Mesiti, F.~Negro, A.~Rucci and F.~Sanfilippo,
Phys. Rev. D \textbf{94} (2016) no.9, 094007
doi:10.1103/PhysRevD.94.094007
[arXiv:1607.08160 [hep-lat]].


\bibitem{Dudal:2017max}
D.~Dudal and S.~Mahapatra,
Phys. Rev. D \textbf{96}, no.12, 126010 (2017)
doi:10.1103/PhysRevD.96.126010
[arXiv:1708.06995 [hep-th]].

\bibitem{Yang:2015aia}
Y.~Yang and P.~H.~Yuan,
JHEP \textbf{12}, 161 (2015)
doi:10.1007/JHEP12(2015)161
[arXiv:1506.05930 [hep-th]].

\bibitem{Chen:2020ath}
X.~Chen, L.~Zhang, D.~Li, D.~Hou and M.~Huang,
[arXiv:2010.14478 [hep-ph]].

\end{thebibliography}
\end{document}